\documentclass[num-refs]{wiley-article}

\usepackage{siunitx}
\usepackage{subcaption}

\papertype{Original Article}
\paperfield{Submitted for peer review to Magnetic Resonance in Medicine}

\title{Split Slice Training Augmentation and Hyperparameter Tuning of RAKI Networks for Simultaneous Multi-Slice Reconstruction}

\abbrevs{RAKI, Robust Artificial-neural-networks for k-space Interpolation; SMS, Simultaneous Multi-Slice; MRI, Magnetic Resonance Imaging.}

\author[1,2]{Andrew S. Nencka, PhD}
\author[2]{Volkan E. Arpinar, PhD}
\author[3]{Sampada Bhave, PhD}
\author[4]{Baolian Yang, PhD}
\author[4]{Suchandrima Banerjee}
\author[5]{Michael McCrea, PhD}
\author[6]{Nikolai J.  Mickevicius}
\author[5,2]{L. Tugan Muftuler, PhD}
\author[1,2]{Kevin M. Koch, PhD}

\affil[1]{Department of Radiology, Medical College of Wisconsin, Milwaukee, WI, 53226, USA}
\affil[2]{Center for Imaging Research, Medical College of Wisconsin, Milwaukee, WI, 53226, USA}
\affil[3]{University of Minnesota, Minneapolis, MN, 55455, USA}
\affil[4]{GE Healthcare, Waukesha, WI, 53188, USA}
\affil[5]{Department of Neurosurgery, Medical College of Wisconsin, Milwaukee, WI, 53226, USA}
\affil[6]{Department of Radiation Oncology, Medical College of Wisconsin, Milwaukee, WI, 53226, USA}

\corraddress{Andrew S. Nencka PhD, Department of Radiology, Medical College of Wisconsin, Milwaukee, WI, 53226, USA}
\corremail{anencka@mcw.edu}

\fundinginfo{GE Healthcare, NINDS U01NS093650, NIA UF1AG051216, USAMRDC W81XWH-12-1-0004}

\runningauthor{Nencka et al.}

\begin{document}

\maketitle

\begin{abstract}
\textbf{Purpose:} Simultaneous multi-slice acquisitions are essential for modern neuroimaging research, enabling high temporal resolution functional neuroimaging and high resolution q-space sampling diffusion acquisitions. Recently, deep learning reconstruction techniques have been introduced for unaliasing these accelerated acquisitions, and Robust Artificial-neural-networks for K-space Interpolation (RAKI) have shown promising preliminary capabilities. The present study systematically examines the impacts of hyperparameter selections for RAKI networks, and introduces a novel training data augmentation formalism. This augmentation approach is analogous to the split-slice formalism used in slice-GRAPPA.  

\noindent\textbf{Methods:} RAKI networks were developed with variable numbers of layers, convolutional filter sizes, numbers of convolutional filters in each layer, numbers of single voxel convolutional filters, batch normalization, training dropout, and split-slice training data augmentation. Each network was trained over a period of five minutes and applied to five different datasets including acquisitions harmonized  with Human Connectome Project protocols. Unaliasing performance was assessed through the L1 norm errors computed between unaliased timeseries and fully-sampled calibration images. 

\noindent\textbf{Results:} Split-slice training data augmentation significantly improved network performance in nearly all hyperparamter configurations. Best unaliasing results were achieved with three layer RAKI networks using at least 64 convolutional filters with receptive fields of 7 voxels, 128 single-voxel filters in the penultimate RAKI layer, batch normalization, and no training dropout with the split-slice augmented training dataset. Networks trained without split-slice augmentation showed symptoms of network over-fitting.

\noindent\textbf{Conclusion:} Split-slice augmentation for simultaneous multi-slice RAKI networks positively impacts network performance. Hyperparameter tuning of such reconstruction networks can lead to further improvements in unaliasing performance.

\end{abstract}

\section{Introduction}
\hspace{\parindent}Magnetic resonance imaging (MRI) has proven to be an exceptionally valuable diagnostic and research imaging platform due to its wide range of available contrast mechanisms and its lack of exposure to ionizing radiation. In the realm of neuroscience research, MRI is a key technology as it can be used to map white matter fiber bundles with diffusion tensor acquisitions \cite{Conturo1999TrackingBrain}, and identify regions of the brain which have changing hemodynamics with cortical activation with blood oxygenation level dependent (BOLD) contrast \cite{Ogawa1990BrainOxygenation}. These acquisitions are based upon ``fast'' imaging techniques, such as echo planar imaging (EPI), which includes a full sampling of Fourier-space (k-space) observations for an image following a single radio frequency (RF) excitation pulse \cite{Mansfield1977Multi-planarEchoes}. Such ``fast'' techniques are essential for acquiring a substantial sets of diffusion weighted images required for diffusion tensor modeling \cite{Alexander2007DiffusionBrain, Lu2006Three-dimensionalImaging, Zhang2012NODDI:Brain, Ozarslan2013MeanMicrostructure}.  Additionally, for functional MRI studies, these methods are required to rapidly capture time series volumetric images with BOLD contrast at reasonable spatial resolution and physiologically relevant time scales with respect to hemodynamic events.

Techniques have been developed to amortize the temporal overhead of the full acquisition processes across multiple 2D slices by simultaneously exciting those slices and de-aliasing the resultant images with  parallel imaging techniques \cite{Larkman2001UseExcited, Moeller2010MultibandFMRI, Jesmanowicz2011Two-AxisAcceleration, Feinberg2010MultiplexedImaging}. Much ongoing development has focused upon improving the parallel imaging techniques for these simultaneous multi-slice (SMS) or multi-band (MB) acquisitions. Improvements include the tagging of each excited slice with unique magnetization phases \cite{Jesmanowicz2011Two-AxisAcceleration, Blaimer2013MultibandMRI}, shifting the apparent locations of the simultaneously excited slices with additional gradient encoding \cite{Setsompop2012Blipped-controlledPenalty}, and parameterizing the de-aliasing algorithm to minimize the leakage of signal between unaliased spatial locations \cite{Cauley2014IntersliceAcquisitions}.

 The significant recent advances in deep convolutional neural networks \cite{Krizhevsky2012ImageNetNetworks} have enabled paradigm shifting approaches to MR image reconstruction. 
A growing number of deep learning techniques for reconstruction are being developed and demonstrated \cite{Hammernik2018LearningData, Han2018DeepMR, Hyun2018DeepReconstruction., Aggarwal2017MoDL:Problems, Biswas2018Model-basedMoDL-STORM, Schlemper2017AReconstruction,Zhu2018ImageLearning,KnollDeepReconstruction}. 



Of specific interest to the present study, convolutional neural networks have been applied to simultaneous multi-slice acquisitions. The Robust Artificial neural networks for K-space Interpolation (RAKI) \cite{Akcakaya2019Scan-specificImaging} approach is a simple, three-layer convolutional neural network is designed to produce and infer dynamically trained neural network models for each imaged subject. It is trained with single band k-space calibration data as the effective training ``label,'' and the aliased slices as the training ``input.'' Thus, the RAKI network is similar to a standard slice-GRAPPA implementation~\cite{slice-GRAPPA}, where the conventional linear computation of the GRAPPA  is replaced by an empirical non-linear fitting algorithm in the convolutional neural network. The initial demonstration of the RAKI network fit models uniquely for each slice and coil~\cite{Akcakaya2019Scan-specificImaging}. More recently, the RAKI algorithm has been modified to simultaneously de-alias multiple slices \cite{Zhang2018AcceleratedNetworks}.

In this work, the concept of ``split-slice'' de-aliasing is introduced to the training data provided to the RAKI network to reduce aliased signal ``leakage" between reconstructed slices. 
Additionally, several variations in network designs and hyperparamters
were analyzed with respect to their impact on unaliasing performance and network training efficiency.  
Network performance was evaluated on human subject timeseries data, whereby the fully-sampled calibration image was used as a reference upon which validation L1 norm loss was utilized as the key performance indicator.

\section{Methods}

\subsection{Network Design}
\begin{figure}[t!]
    \centering
    \includegraphics[width=\textwidth]{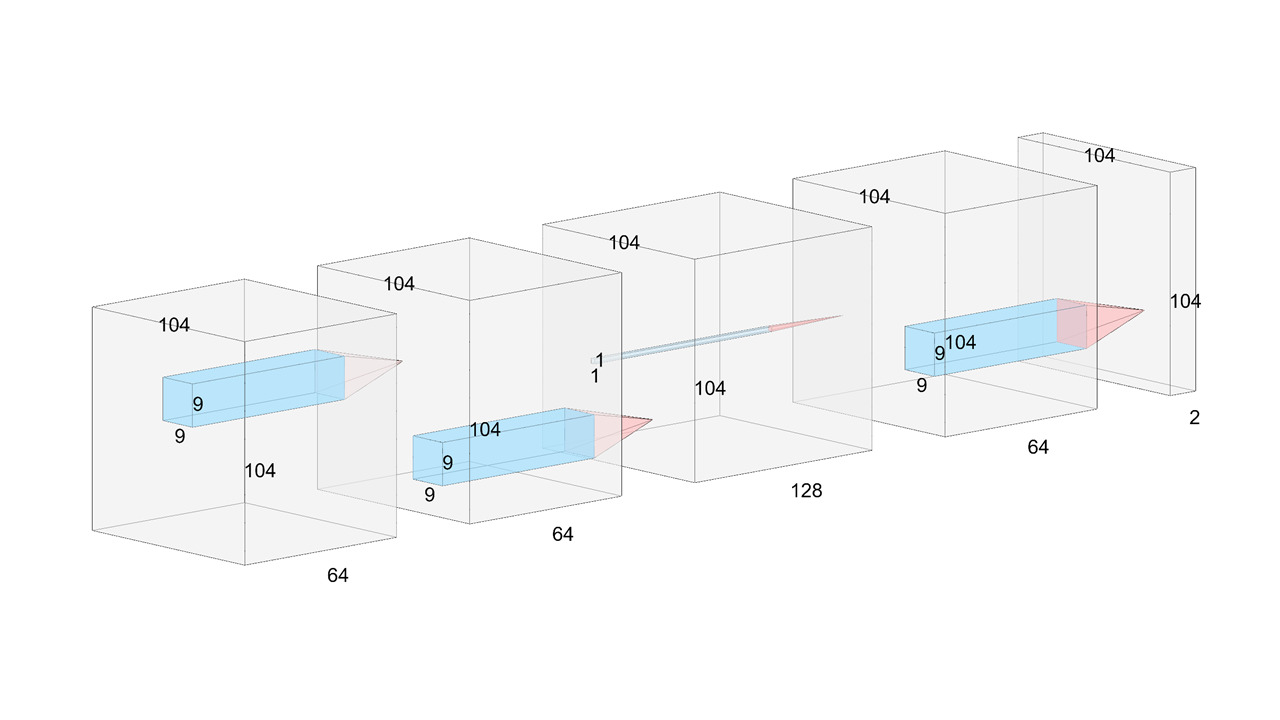}
    \caption{Example RAKI network. The input is a 104x104 k-space array, with real- and imaginary-valued layers for each of 32 coils, yielding 64 input layers. In this example case, the network has 4 convolutional layers, with the penultimate layer including 128 unit convolutions, and all other layers including 64 convolutions with 9x9 kernels. The output yields a full k-space array for the real- and imaginary-channels of one reconstructed coil.}
    \label{fig:NetExample}
\end{figure}

\hspace{\parindent}A commonly-utilized network architecture, based upon the previously published RAKI network \cite{Akcakaya2019Scan-specificImaging}, was selected as the primary network architecture for the current study . Figure \ref{fig:NetExample} shows an example RAKI network, wherein the multitude of network hyperparameters are illustrated. Apparent in the figure are 4 convolutional layers, 9$\times$9 convolutional filter size, 64 convolutional filters for each layer, and 128 convolutional filters in the penultimate layer. A ReLU activation function is applied to all convolutions except for the final convolution \cite{Nair2010RectifiedMachines}. 

To assess the impact of network hyperparameters on training efficiency and unaliasing performance, various components of the RAKI network were independently varied. Hyperparameters examined in this work include: number of convolutional layers, convolutional filter size, number of convolutional filters in each layer, number of convolutional filters in the penultimate layer with unity filter size, inclusion of batch normalization following convolutions, and inclusion of 50\% dropout following convolutions.  When dropout and batch normalization were applied, they were applied following all convolutions except the final convolution. Table \ref{tab:Hyperparameters} shows the range of hyperparameters tested for each metric. All hyperparameters were varied independently in a grid search.

\begin{table}[]
    \centering
    \begin{small}
    \begin{tabular}{|l|c|c|c|c|c|c|c|c|c|}
         \hline
         & Number & Filter  & Number  & Penultimate & Batch         & Dropout   &  Split-Slice\\
         & Layers & Size    & Filters &  Filters     & Norm. &          & Augmentation\\
         \hline
  Values & 1:2:7 & 1:2:11 & 32:32:128   & 64,128:128:512 & T/F & T/F  & T/F\\
        \hline
        Total & 4 & 6 & 4 & 5 & 2 & 2 & 2 \\
        \hline
    \end{tabular}
    \end{small}
    \caption{Hyperparameters varied in the evaluation of RAKI networks for slice unaliasing. With a grid search, this results in 3072 unique hyperparameter combinations that were used to train networks for each dataset.}
    \label{tab:Hyperparameters}
\end{table}

The neural networks in this study were developed using PyTorch 1.3.1 \cite{PaszkeAutomaticPyTorch}, with open source code available at \cite{RakiNetwork/README.mdNenckaLab/RakiNetwork}. For consistent comparison, a subset of training hyperparameters were held constant for this investigation, with appropriate values identified though a limited parameter space search. Bias terms in convolutions were set to zero. Learning rate was set to 0.0001, and the ADAM optimizer was used ($\beta$s 0.9 and 0.999; weight decay 0) \cite{Kingma2015Adam:Optimization}. L1 normalized loss was utilized for the optimization cost function. Convolutions were performed with only one group, allowing real and imaginary observations in each layer to contribute to real and imaginary observations in the following layer. Convolutional layers included a zero padding equal to half the convolutional kernel size to yield output arrays matching the dimensionality of input arrays. The random number generator was seeded with the value of 42. In place of setting the number of epochs to be a constant, training time was fixed at 5 minutes for each network. For split-slice training (described in the next section), batch size was set to 48, while standard RAKI includes only one training set, making batch size irrelevant for training.

Training was performed with hardware acceleration using graphical processing units. Each network was trained using one NVIDIA K80 processor, and the 3072 unique networks were trained for each dataset in parallel utilizing a cluster including 24 of these graphical processing units. Training input data was either selected as the first time point in the EPI time series following the auto-calibration block, or was generated through a synthetic aliasing as described below, and training output data were selected as the Fourier unaliased auto-calibration data described below. 

\subsection{Split-Slice Training Augmentation}
\hspace{\parindent}The use of deep neural networks in k-space for image unaliasing can be viewed as a non-linear implementation of the GRAPPA technique \cite{Griswold2002GeneralizedGRAPPA}. In the present context, the GRAPPA kernel, which is conventionally estimated using methods of linear regression, is replaced by a convolutional neural network. In fact, a degenerate RAKI network, including one convolutional layer with one filter and without activations is equivalent to a conventional GRAPPA implementation.  

Within the published RAKI formalism, a unique set of kernels, or neural network weights, are computed for each coil and each slice from a single aliased input k-space array across all coils \cite{Akcakaya2019Scan-specificImaging}. For split-slice GRAPPA, however, individual k-space observations within a slice from all coils are supplied as inputs.  Multi-slice k-space estimates are then generated as outputs, such that only the observations corresponding to the input slice location are non-zero for the desired slice and coil \cite{Cauley2014IntersliceAcquisitions}.

\begin{figure}[t!]
    \centering
    \includegraphics[width=\textwidth]{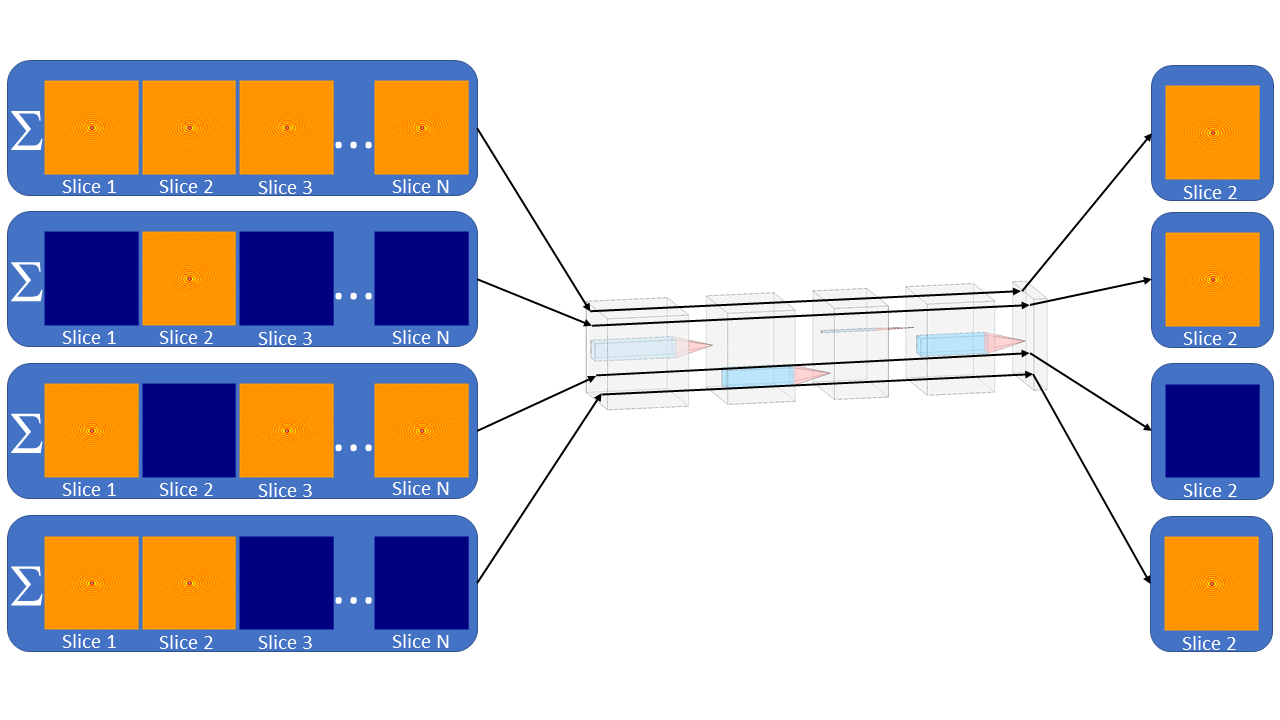}
    \caption{Split slice training augmentation. Aliased k-space data, summed across slices, is input into the network on the left, and the RAKI network for unaliasing slice 2 is trained with expected output arrays shown on the right. Each of the four rows corresponds to a unique input/output set for training. RAKI, as originally described, is trained with fully aliased multi-coil k-space as input and an unaliased k-space for one slice and coil combination as output (first row). Training data can be augmented by creating synthetically aliased k-space data by summing the k-space observations for subsets of slices in the excited packet. When the slice to be unaliased with the RAKI network is not included in the synthetically aliased k-space, the network is trained with an output of zero (third row).}
    \label{fig:SplitSlice}
\end{figure}

Analogous to split-slice GRAPPA, single band k-space calibration data can be utilized for training in a split-slice RAKI technique. A set of training data is built by summing subsets of single slice k-space observations from the single-band calibration data for each slice in the packet excited with the SMS acquisition. A graphical representation of this yielding four training sets is shown in Fig. \ref{fig:SplitSlice}. The first set in the figure includes the traditional RAKI training dataset, wherein the fully aliased k-space for all coils is input and the k-space for the coil and slice from the network is output. In subsequent rows of the figure, different subsets of slices are summed for each coil and the k-space for the network's target slice and coil is output. If the target slice k-space is not included in the synthetically aliased input (i.e. zero-valued), then the training output is also set to zero. Each slice of the packet of aliased slices can either be included or not included in the synthetically aliased “input” dataset to the neural network. With that binary decision logic, this yields 2$^n$ unique combinations of subsets of slices in the packet of $n$ simultaneously excited slices that can be synthetically aliased. As such, this training data augmentation technique yields $2^n$ sets of training data for an $n$-fold SMS acceleration.

In this work, each network was trained with the original RAKI training data set as well as the augmented split-slice training data set. With the 8-fold SMS acceleration, this yields 256 unique training datasets, compared to the single training dataset used in RAKI.

\subsection{Data Acquisition}
\hspace{\parindent}This study was approved by the local Institutional Review Board and prospective written consents were obtained from all participants. Experiments were conducted on a GE Healthcare Discovery MR750 3.0T system (GE Healthcare, Waukesha WI), using a 32-channel head receive coil (Nova Medical, Wilmington MA). Data from five research participants, enrolled in a larger study of sport concussion \cite{Koch2018QuantitativeConcussion}, were preserved for off-line reconstruction. Imaging parameters included: 30 ms TE, 800 ms TR, 503 repetitions, 50$^\circ$ flip angle, 104$\times$104 acquisition matrix, 20.8 cm field of view, 2 mm thick slices, 8-band SMS acceleration, $\frac{1}{3}$ CAIPI field of view shift, and 64 total slices, harmonized with the Human Connectome Project acquistion protocol \cite{Smith2013Resting-stateProject}. Calibration images were generated using an auto-calibration technique in which the first 16 repetitions of the time series included a modulation of the RF excitation phase of each slice through a Fourier encoding scheme \cite{Ferrazzi2019AutocalibratedMapping}. Unaliased k-space calibration data were generated through the Fourier transform of those first 16 calibration repetitions.

\subsection{Network Performance Assessment}
\hspace{\parindent}Following network training, each trained model for a slice and coil combination for a given acquisition was used to unalias 20 unique time points in the EPI time series that were not included in training set. Those 20 time points were equally spaced throughout the time series acquisition. The performance of this inference was measured using the same L1 loss function used in training, and was reported as an average for each real or imaginary voxel value across all voxels and testing time points.

Along with L1 error performance, computational efficiency was also characterized for each of the implemented training approaches.  As training performance is likely to improve with number of epochs completed, and a more computationally efficient network will complete more epochs in the given training time than a more computationally intensive network. Additionally, the use of GPU resources were measured following training of each model, by using the vendor provided ``NV-SMI'' program to report the maximum and average GPU processor and memory usage. A more computationally efficient network will maximize utilization of GPU resources.  While computational performance  metrics hold little direct bearing on unaliasing performance, it is anticipated that models which are more computationally efficient will maximize the number of training epochs completed in a fixed period of time, thereby improving performance.

To compare the relative performance ranking of each network across acquired subjects, a subject normalized loss function was computed. This normalized loss was the ratio of the network's loss function computed on the testing dataset to the minimal testing loss function for that subject across all 3072 networks. With this normalized loss function, all networks in each subject were ranked. An optimal network design should yield results which are in the highest percentiles across all subjects. Also, a higher percentile that yields an optimal network with a consistent set of hyperparameters across subjects should be indicative of the robustness of a training dataset parameterization.

\section{Results}
\begin{figure}[t!]
    \includegraphics[width=.98 \textwidth]{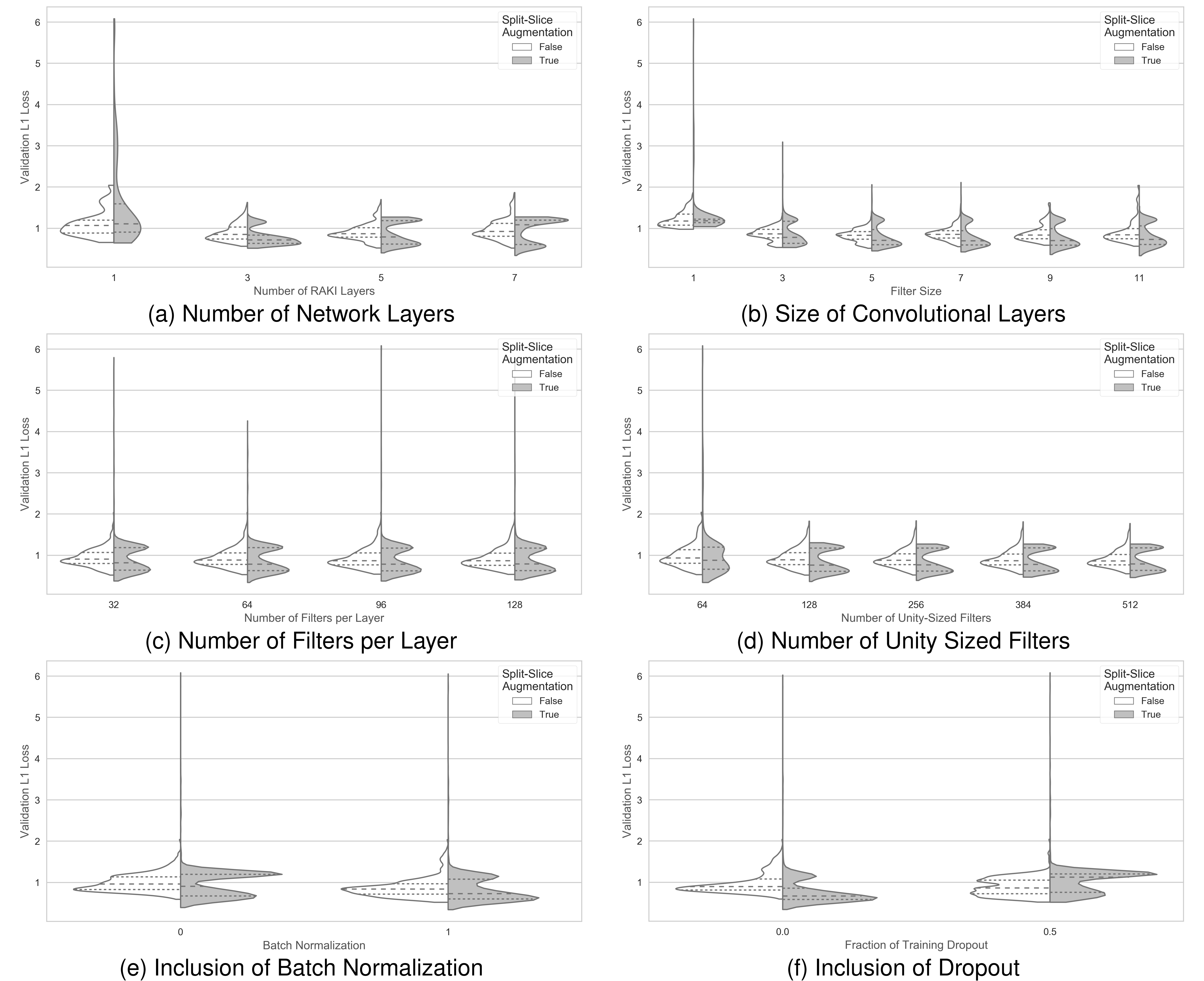}
    
    
    
    \caption{Violin plots of validation dataset loss functions with and without split-slice training data augmentation for the considered network hyperparameters. Median loss function values (long dashes), and the 25$^{th}$ and 75$^{th}$ percentiles (short dashes) are shown in the violin plots.}
    \label{fig:resultPlots}
\end{figure}

\hspace{\parindent}
The performance impact from the seven evaluated network design hyperparameters 
are presented in detail in the following sections. Values were compared, unless otherwise noted, with non-parametric Wilcoxon signed rank tests \cite{Wilcoxon1945IndividualMethods}. In scenarios where multiple parameters were analyzed using linear regression, slopes are reported with p-values for a hypothesis test of the slope being non-zero. Plots of validation L1 loss as a function of hyperparameter variations are shown in Figure \ref{fig:resultPlots}. Similar plots showing numbers of epochs completed, GPU processor usage, and GPU memory usage are presented as supplemental Figures (Figs. \ref{fig:epochSSPlots}-\ref{fig:gpuPlots}).

\subsection{Split-Slice Augmentation}
\hspace{\parindent}Split-slice augmentation of training data improves the performance of trained RAKI networks. The split slice augmentation yielded a median reduction in loss function value of 0.092 (paired t-test p<1e-8). In the full group of tested models, 60\% of models perform better with the split slice augmentation, and in the subset of best performing networks wherein both the networks trained with standard RAKI and split slice RAKI yield validation cost functions of less than unity, 81\% of models perform better with the split splice augmentation. In Figure \ref{fig:resultPlots}, the improved performance of split-slice training is apparent as the median cost function values computed on the validation training sets (dashed line) is lower in the split-slice augmented training set (gray) compared to the traditional RAKI training set (white). 

Due to the fact that an epoch with split slice training includes 256 times the number of training datasets compared to standard RAKI training, a number of significant differences were noted in the training process. The number of epochs completed with the split-slice augmentation was dramatically less compared to standard training, with a median number of epochs completed of 112 versus 11,737 for the two methods. Similarly, with increased training data volume, the split-slice method utilized more GPU memory, with a median usage of 40\% of the available GPU memory across all networks, compared to 34\% with standard RAKI. GPU processing utilization, however, was found to be very similar between the two training techniques, with median usage of 89\% and 88\% for split-slice and standard training, respectively.

\subsection{Number of Layers}
\hspace{\parindent}The number of layers in the RAKI network was found to have a strong impact on unaliasing performance. Results are shown in Figure \ref{fig:resultPlots}a. On average and with both conventional/spit-slice training sets, a network architecture of 3 layers was found to yield lowest validation loss.  Markedly degraded performance was observed in the case of a single-layer network, with an increase in cost function loss of 0.75 (p<<1e-10) with split slice training and 0.22 (p<<1e-10) with traditional training. Less marked, but still significant, increases in cost function loss less than or equal to 0.2 (p<1e-3) were present when there were 5 or 7 layers.

Increasing the number of layers in a RAKI network leads to an increase in the number of model weights. With increased model complexity, fewer epochs were run during the controlled training time. With split-slice training, a strong linear trend was observed (p<<1e-10) with nearly 7.8 fewer epochs completed for each additional layer introduced, and a similar trend was observed (p<<1e-10) with standard training with 5,000 fewer epochs completed with each additional layer. Decreased training epochs in the controlled training time could account for decreased model performance with increases in the number of layers. Unsurprisingly, increased layers corresponded to increased GPU memory usage. Additionally, increasing from one to three layers increased GPU usage from 65\% to 81\%, and GPU usage remained saturated around 81\% with further increases in the number of layers as the GPU usage was saturated.

\subsection{Filter Size}
\hspace{\parindent}Filter size was found to have different relationships with network performance based upon the training set which was used, as can be seen in Figure \ref{fig:resultPlots}b. With split slice training, a trend for increased performance with kernel size was present (slope -0.041, p<<1e-10), while a trend showing a reduced impact of kernel size was present with standard RAKI training (slope -0.022, p<<1e-10). In spite of these linear trends, the lowest validation set loss functions were observed with a filter sizes of 7 and 9 in the split slice training set and filter size of 5 with the standard training set.

Increasing the filter size in a RAKI network leads to an increase in the number of model weights, and fewer epochs were run during the training time. With split-slice training, a linear trend was observed (p<<1e-10) with nearly 8.1 fewer epochs completed for each added point of filter width, and a similar trend was observed (p<<1e-10) with standard training with 3300 fewer epochs completed with each additional point of filter width. Increased filter sizes used more GPU processor cycles, with filter sizes from one to five utilizing 71-75\% of the clock cycles and larger filters saturating the GPU with 83-94\% usage. In each case, an increase of 2\% GPU usage was correlated with each additional point of filter width (p<<1e-10). Memory usage with split-slice training remained relatively constant across filter sizes at approximately 31-46\%, with increases in usage associated with more filters (slopes of 0.72\% per point and 1.2\% per point [p<<1e-10] for split-slice and standard training, respectively).

\subsection{Number of Filters}
\hspace{\parindent} As shown in Figure \ref{fig:resultPlots}c, the number of filters yielded minimal impact on the performance of the networks tested in this study. No significant trends were observed with split-slice augmentation. A very slight trend of improvement with more filters was observed with standard training, with a loss function reduction of 0.00018 with each added filter (p<0.03). 

Increasing the number of filters in a RAKI network leads to an increase in the number of model weights, similar to increasing the number of layers. With more filters, fewer epochs were run during the training time. With split-slice training, a linear trend was observed (p<<1e-10) with nearly 0.14 fewer epochs completed for each added filter, and a similar trend was observed (p<1e-5) with standard training with 27 fewer epochs completed with each additional filter. Increased numbers of filters minimally impacted GPU processor cycle usage, with approximately 75\% to 83\% GPU usage with both training sets. There was no significant trend in GPU usage with split-slice training, and a modest trend of reduced GPU usage by 0.022\% for each filter (p<0.01) with standard traing. Similarly, modest increases in GPU memory utilization were only observed with standard RAKI training with increased number of filters (0.053 increase of percentage used for each additional filter, p<1e-10).

\subsection{Penultimate Layer Filters}
\hspace{\parindent}The penultimate layer of the RAKI network includes a convolution with filters corresponding to individual voxels. This layer exists only in RAKI networks wherein the number of layers is greater than or equal to two. As seen in Figure \ref{fig:resultPlots}d, the number of filters in this layer offers minimal impact on unaliasing performance when at least 128 filters are used with the split-slice training data. Best performance with the split slice data is achieved with 128 filters, which is significantly better than 64 filters (p<1e-9), 384 filters (p<0.02), and 512 filters (p<0.002), while offering a non-significant improvement compared to 256 filters. With standard RAKI training, best results are achieved with 512 filters, with only significant improvement over 128 filters (p<0.02) and 64 filters (p<1e-10).

Increasing the number of filters in the penultimate RAKI network layer leads yields similar results as with increasing the number of filters in the other RAKI network layers. With more filters, fewer epochs were run during the training time. With split-slice training, a linear trend was observed (p<<1e-10) with nearly 0.10 fewer epochs completed for each added filter, and a similar trend was observed (p<<1e-10) with standard training with 32 fewer epochs completed with each additional filter. Linear trends of increasing GPU usage with increasing number of filters were once again observed, with slopes of 0.012 (p<1e-10) and 0.0069 (p<1e-4) for split-slice and standard training. With both training sets, GPU memory usage was reduced when 64 filters were used (32\% and 31\% with split-slice and standard training, respectively), while it remained stable at 38-41\% and 35-37\% for split-slice and standard networks with other numbers of filters in this layer.

\subsection{Batch Normalization}
\hspace{\parindent} Inclusion of batch normalization yields improved results. This technique, which addresses internal covariate shifts, has been previously shown to improve training convergence \cite{Ioffe2015BatchShift}. In the networks utilized in this work, batch normalization reduces the median validation loss function value with standard training by 0.12 (p<<1e-10) and split-sliced augmented training by 0.17 (p<<1e-10). This marked improvement is seen in Figure \ref{fig:resultPlots}e.

Including batch normalization increases the computations performed in the deep neural network. As such, the median number of epochs completed was reduced with batch normalization by 3.2 (p<0.001) and 2700 (p<1e-7) epochs with split-slice and standard training, respectively. GPU usage increased from 89\% to 94\% (p<0.01) with split-slice training with batch normalization, and decreased from 93\% to 89\% (p<0.01) with standard training. Similarly GPU memory usage increased from 36\% to 39\% (p<1e-6) with split-slice training, and decreased from 32\% to 31\% (p<0.001) with standard training.

\subsection{Dropout}
\hspace{\parindent} Dropout layers are included in the training of networks to reduce the probability of over-fitting during training \cite{GeronHands-onSystems}. With a very limited training set wheb using standard RAKI training\cite{Akcakaya2019Scan-specificImaging}, over-fitting is of great concern, and the inclusion of 50\% dropout layers yielded significantly improved network performance, with a decrease in validation loss computation of 0.089, as shown in Figure \ref{fig:resultPlots}f (p<<1e-10). Conversely, with split slice augmentation and the limited training duration of 300 seconds, the inclusion of 50\% dropout layers yields decreased network performance, with an increase of the validation loss function of 0.24 (p<<1e-10).

Dropout reduces the number of neurons fit in the deep network in each batch. Including dropout did not yield a difference in the number of epochs run during the training period with either training set. However, including dropout led to increased GPU usage 89\% to 94\% (p<0.001) and 87\% to 95\% (p<1e-10) with split-slice and standard training. Including dropout did not change the memory usage of split-slice training, although it increased memory usage from 29\% to 34\% (p<1e-10).

\begin{figure}[t!]
    \centering
    \includegraphics[width=\textwidth]{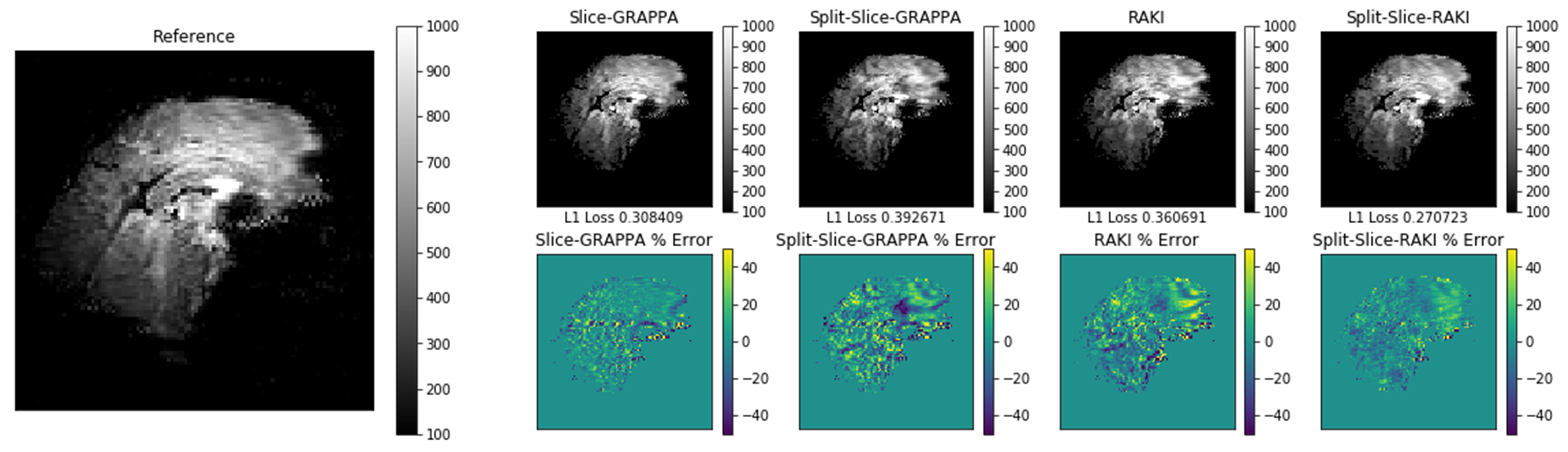}
    \caption{Representative images arising from the RAKI networks, as well as from traditional GRAPPA and split-slice-GRAPPA algorithms. The utilized reference image is shown on the left, unaliased images are shown in the top row on the right, and images of percent error between the unaliased images and the reference image are shown on the bottom row on the right. Below each unaliased image, the k-space L1 norm error (reported as the primary performance metric with deep learning techniques) is listed for reference.}
    \label{fig:GrappaComp}
\end{figure}

\subsection{Best Performing Networks}
Different sets of optimal parameters were identified with split-slice and standard RAKI training. With split-slice training, an optimal network was identified to include: batch normalization, no training dropout, 64 unity-sized filters, 64 convolutional filters with 9$\times$9 receptive fields, and 7 RAKI network layers. This parameterization was in the 99.25$^{th}$ percentile across all five subjects. With standard training, the optimal identified network included: batch normalization, training dropout, 128 unity-sized filters, 64 convolutional filters with 5$\times$5 receptive fields, and 5 RAKI network layers. With standard training, best performing network architectures were less stable across subjects, and this network was in the 97$^{th}$ percentile across all five subjects. 

Resulting images for a representative slice, coil, and subject from these networks, as well as from standard GRAPPA and split-slice-GRAPPA implementations, are shown in Fig. \ref{fig:GrappaComp}. Percent error maps show obvious differences in performance between the methods, with most notable differences in regions in the frontal lobe. Also noted in Fig. \ref{fig:GrappaComp} are the L1 loss function values calculated for the underlying k-space data used to reconstruct the unaliased images. The k-space L1 loss function values are consistent with the image-space error maps, and show that split-slice RAKI and standard GRAPPA yield images with reduced error compared to standard RAKI and split-slice GRAPPA.

\section{Discussion}
\hspace{\parindent}
There are two primary results of this study. First, the newly presented construct of split-slice training data augmentation improves RAKI network performance.  Second, RAKI network hyperparameters significantly impact unaliasing performance and should be tuned for robust applications in light of the selected training dataset.

The concept of split-slice training data augmentation was shown to be beneficial for the application of trained networks to acquired data which was not included in the training data set. Because RAKI networks need to be trained on a subject-by-subject basis, the implementation of these networks are limited by the requisite training time.  Significant over-fitting challenges are present in non-augmented training data sets, which is only partially mitigated with  the addition of dropout layers. However, adding diversity to the training data set with split-slice augmentation was shown to substantially reduce the confounds of over-fitting.  This suggests that training data augmentation with the split-slice formalism is preferred over training dropout in RAKI networks.

The result of this study can be used to provide a recommended set of RAKI network hyperparameters for human connectome project harmonized acquisitions when using the common experimental setup (pulse sequence, MRI system model, and radio frequency receive coil) which was used in this work. Within the context of this study, the best RAKI results arise from networks with: split slice training augmentation, seven layers, 64 convolutional filters with 9$\times$9 voxel receptive fields, 64 single voxel convolutional filters in the penultimate layer, and the inclusion of batch normalization. If the number of training epochs of the split-slice augmented data are kept low, on the order of 112 as were performed in 5 minutes in the presented work, the inclusion of dropout layers to reduce over-fitting is not advised as it decreases the inference performance of the fit networks. 

The performance dependence of RAKI networks on filter size was found to rely on the presence of split-slice training augmentation. With standard RAKI training, wherein only one dataset is used to fit the model weights, networks fit with larger kernel sizes performed poorly when applied to the validation data. This could be indicative of network over-fitting. By increasing the number of variables fit in the deep neural network by increasing filter size and including a very limited training data set, there were more degrees of freedom in the fitting procedure to yield an over-fit model. Conversely, with training data augmentation through the split-slice technique, the probability of over-fitting was reduced because there was more diversity in the training data set. As such, the larger kernel sizes were found to perform better with split-slice training.

On the other extreme, the simplest of the evaluated network designs (fewer layers and filters) were found to have poorer performance than networks with more elaborate architectures. In the most extreme case, a single-layer network with a single filter, which is equivalent to slice-GRAPPA, yielded performance outside of the 25$^th$ percentile across all networks for each of the considered subjects. This suggests that the RAKI networks perform better than simple slice-GRAPPA algorithms. However, this work was not designed to robustly compare these algorithms because GRAPPA algorithms include a vast number of parameters which can, in turn, be optimized \cite{Muftuler2020OptimizationReconstruction}. Future work to objectively compare optimal slice-GRAPPA and RAKI-based algorithms, though beyond the scope of this work, are warranted.

The violin plots provided in this analysis show a multi-modal distribution that is more apparent in the networks trained with the split-slice formalism.  When the loss functions are normalized within subjects, this multi-modal loss appearance fades. This observation suggests that data from some subjects systematically yields images with sub-par unaliasing and warrants further analysis in larger study cohorts.  


While the NVIDIA K80 GPUs include chips (``Kepler'' generation) which are several generations behind the current state of the art, their performance-per-dollar make them well suited for highly-parallelized deployments. A subset of networks fit in this work were additionally re-fit using a current generation NVIDIA Titan V GPU. This newer ``Volta'' generation includes specific ``Tensor Cores'' as well as more CUDA cores and higher clock and memory speeds to speed the process of deep learning network training. It was found that the ``Volta'' chips performed approximately 150\% more training epochs in the same 5 minute training time. As such, if the presented work were run on a lower tier of current generation hardware, the presented results are expected to be achievable in about 3 minutes 20 seconds of training in place of 5 minutes of training.

Training time in this work was limited to 5 minutes for each network for the reconstruction hardware utilized in this study. It is well known that increasing training epochs yields improved network performance \cite{GeronHands-onSystems}. However, the RAKI network needs to be trained for each imaging exam, and a unique network needs to be trained for each slice and coil. This computational limitation, therefore, leads to a limitation in the application of the RAKI network: practical applications are limited by the available computational resources and the amount of time which is allowed for network training. It is expected that the performance of the tested RAKI networks could be improved if training were to be performed with an increased number of training epochs, and as described above, improvements in computational hardware can enable such an improvement. Further, it is reasonable to expect that transfer learning, in which RAKI networks are trained using diverse training sets over durations orders of magnitude greater than the 5 minutes are used to identify initial network weights for more abbreviated exam-specific training. It is expected that, for exam-specific RAKI training to be deployable on a production scale, the deployment of such techniques to reduce training time will be essential.

In this study, the split-slice formalism was introduced as a means to augment training data, thereby improving unaliasing performance with aliased observations that were not included in the training dataset. That improvement was observed in this presented work. With GRAPPA, the split-slice formalism was introduced to reduce slice leakage, or cross-talk, following unaliasing. This work does not include an analysis of slice cross talk or applications to functional neuroimaging or diffusion acquisitions. With recommended network architectures identified through this investigation of deep network hyperparameters, the evaluation of these networks with such acquisitions will be the subject of future studies.

The acquisitions utilized for this study leveraged human connectome project harmonized protocols. Specifically, networks were considered for only 8 packets of 8-fold SMS accelerated slices in the axial plane with 2 mm isotropic resolution and acquisition utilizing a 32-channel Nova Medical head coil. Analyses seeking to optimize hyperparameters associated with conventional slice unaliasing methods have shown that there is a dependence of optimal hyperparameters based upon acquisition and acceleration parameters \cite{Muftuler2020OptimizationReconstruction}. As such, care should be taken with respect to generalizing the results of this work to other acquisition parameterizations. However, as shown with the aforementioned investigation of conventional unaliasing approaches, it is expected that optimal hyperparameters for human connectome project harmonized acquisitions will be among the better performing hyperparameters for other, less aggressive acceleration factors.


\section{Conclusion}
Split-slice training data augmentation yields improved unaliasing performance when deploying RAKI deep neural networks for slice-unaliasing.   In addition, increasing the number of network layer filters can further improve unaliasing performance of these networks at the cost of increased computational time.

\bibliography{References}


\begin{figure}[p!]
    \includegraphics[width=.98 \textwidth]{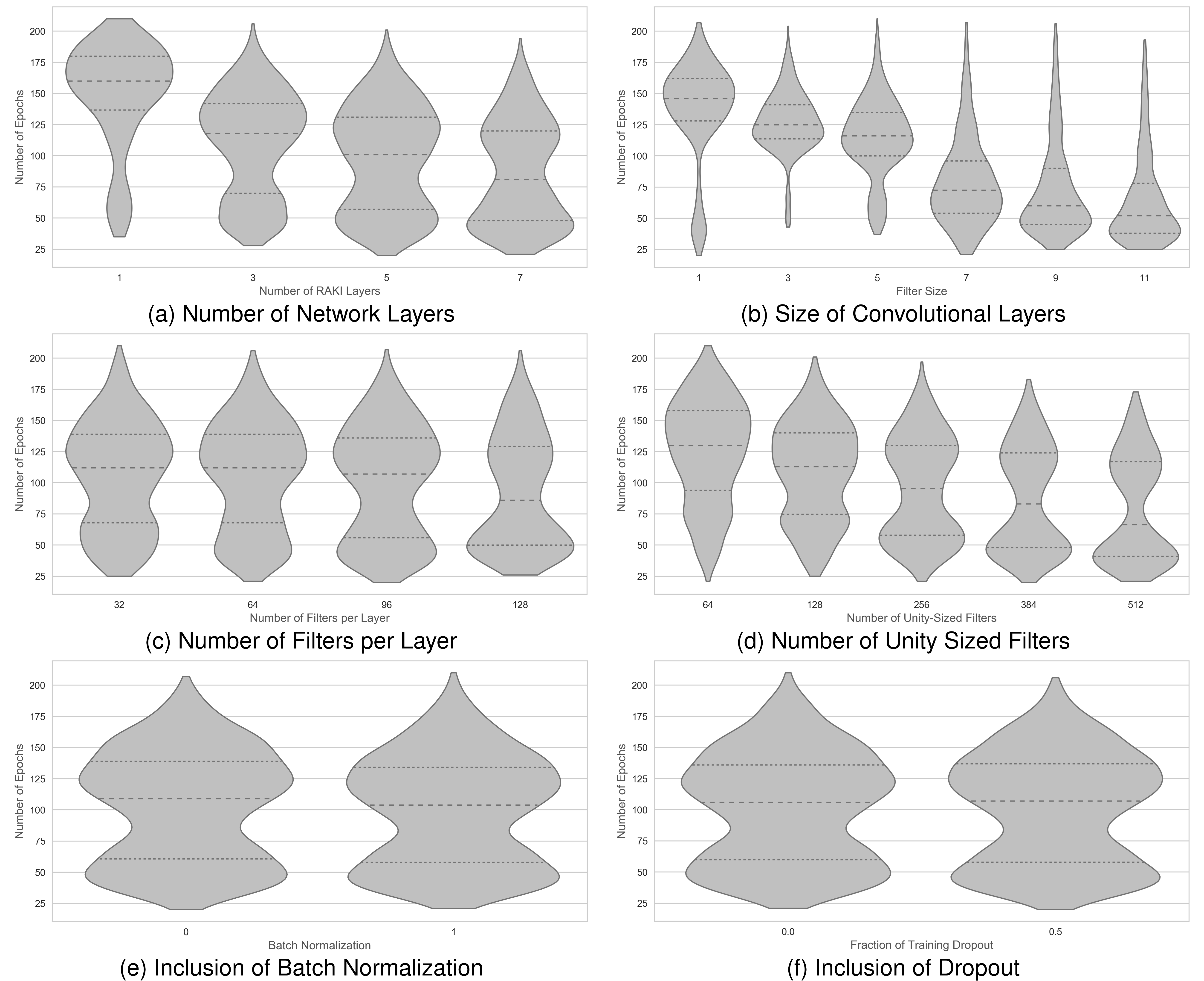}
    
    
    
    \caption{Violin plots of epochs completed with split-slice training data augmentation for the considered network hyperparameters.}
    \label{fig:epochSSPlots}
\end{figure}

\begin{figure}[p!]
    \includegraphics[width=.98 \textwidth]{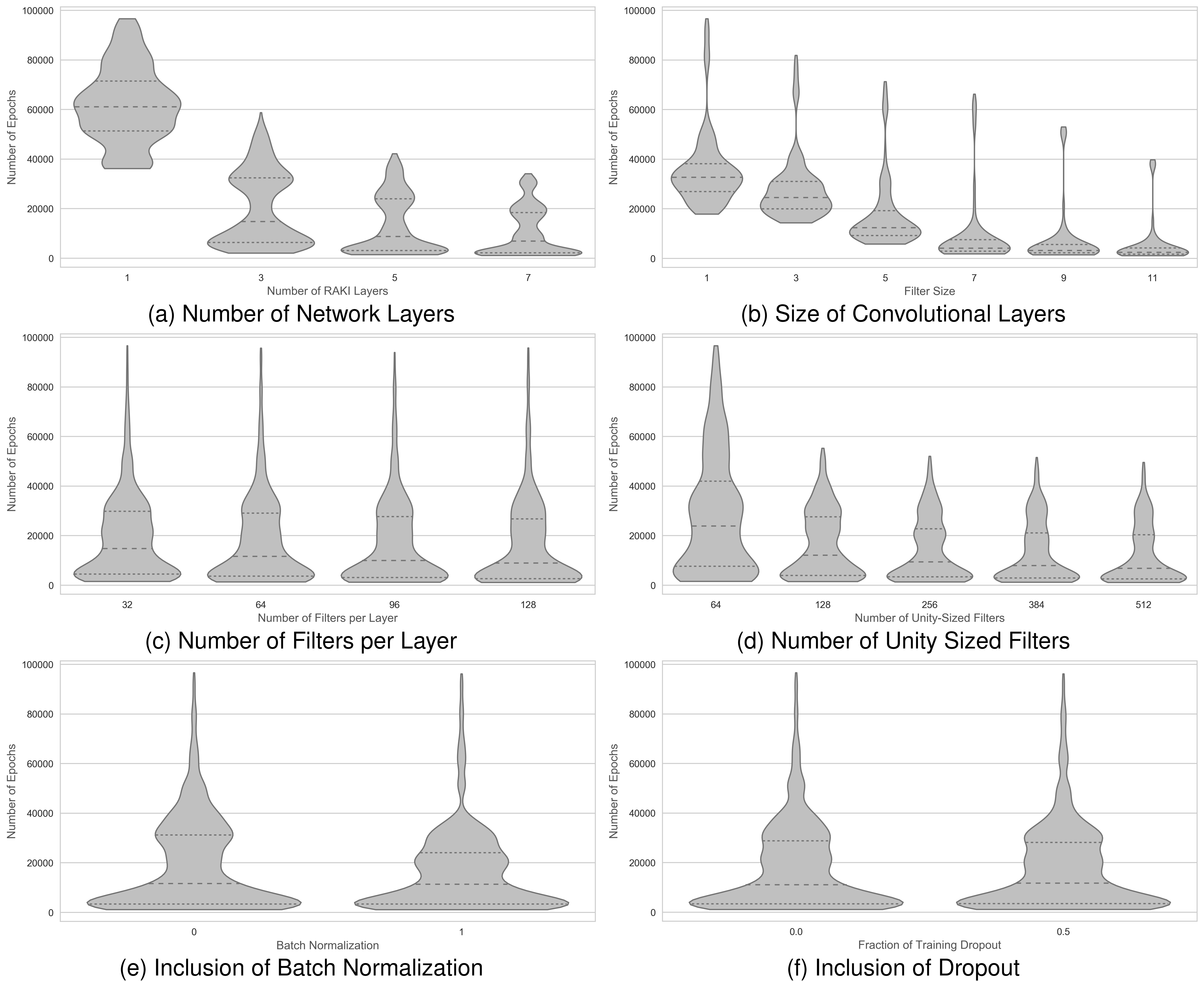}
    
    
    
    \caption{Violin plots of epochs completed with standard training data augmentation for the considered network hyperparameters.}
    \label{fig:epochSSPlots}
\end{figure}

\begin{figure}[p!]
    \includegraphics[width=.98 \textwidth]{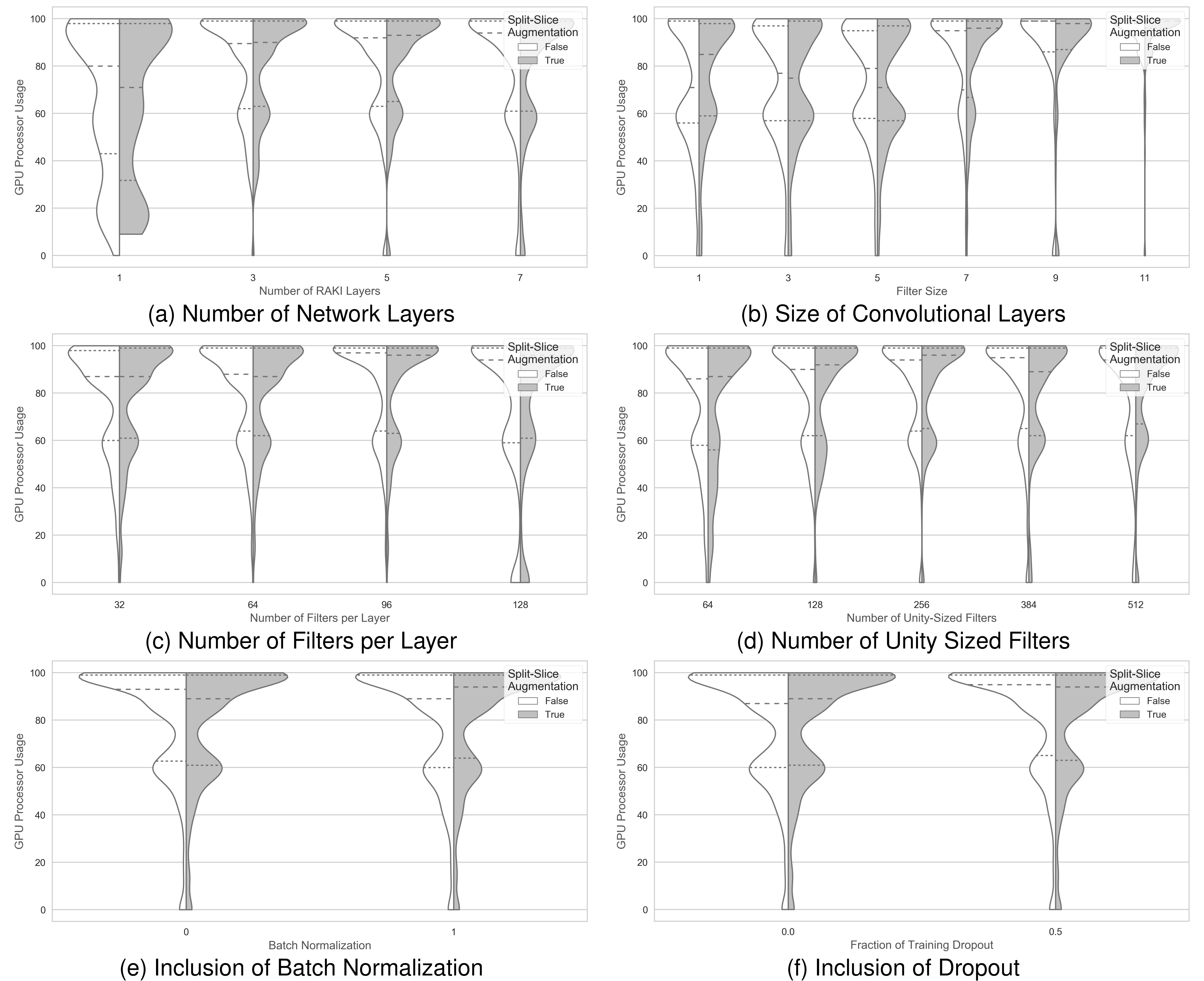}
    
    
    
    \caption{Violin plots of GPU processor usage with and without split-slice training data augmentation for the considered network hyperparameters.}
    \label{fig:gpuPlots}
\end{figure}

\begin{figure}[p!]
    \includegraphics[width=.98 \textwidth]{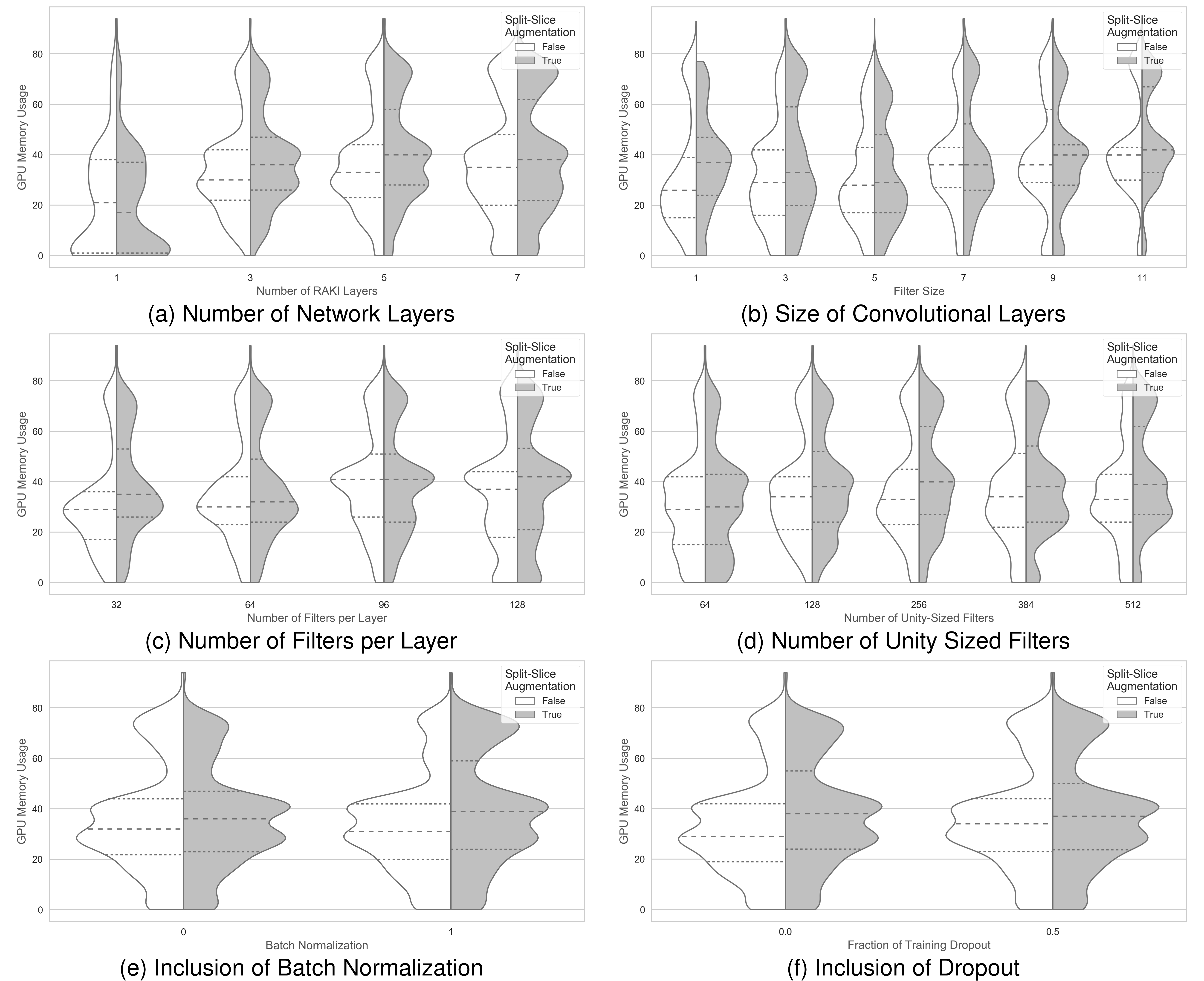}
    
    
    
    \caption{Violin plots of GPU memory usage with and without split-slice training data augmentation for the considered network hyperparameters.}
    \label{fig:gpuPlots}
\end{figure}

\end{document}